\documentclass[12pt,preprint]{aastex}
\usepackage{lscape}
\newcommand{\lya}{Ly$\alpha$}
\newcommand{\kms}{km~s$^{-1}$}
\newcommand{\dnzcg}{$dN/dz|_{\rm GRB}^{\rm CIV}$}
\newcommand{\dnzcq}{$dN/dz|_{\rm QSO}^{\rm CIV}$}
\newcommand{\mdnzcq}{dN/dz|_{\rm QSO}^{\rm CIV}}
\newcommand{\mdnzcg}{dN/dz|_{\rm GRB}^{\rm CIV}}
\newcommand{\dnzmgg}{$dN/dz|_{\rm GRB}^{\rm MgII}$}
\newcommand{\mdnzmgg}{dN/dz|_{\rm GRB}^{\rm MgII}}
\newcommand{\dnzmgq}{$dN/dz|_{\rm QSO}^{\rm MgII}$}
\newcommand{\mdnzmgq}{dN/dz|_{\rm QSO}^{\rm MgII}}

\shorttitle{\ion{C}{4} in GRB sightlines}
\shortauthors{Tejos, N. et al.}

\begin{document}

%% LaTeX will automatically break titles if they run longer than
%% one line. However, you may use \\ to force a line break if
%% you desire.

\title{On the Incidence of \ion{C}{4} Absorbers Along the Sightlines to 
Gamma-Ray Bursts}

%% Use \author, \affil, and the \and command to format
%% author and affiliation information.
%% Note that \email has replaced the old \authoremail command
%% from AASTeX v4.0. You can use \email to mark an email address
%% anywhere in the paper, not just in the front matter.
%% As in the title, use \\ to force line breaks.

%JXP -- I think the following will be the full author list.}

\author{Nicolas Tejos\altaffilmark{1}, 
        Sebastian Lopez\altaffilmark{1}, 
        Jason X. Prochaska\altaffilmark{2}, 
        Hsiao-Wen Chen\altaffilmark{3}, 
	Miroslava Dessauges-Zavadsky\altaffilmark{4}
}
%\email{ntejos@das.uchile.cl}

\altaffiltext{1}{Departamento de Astronom\'{\i}a; Universidad de Chile; Casilla 36-D, Santiago, Chile; ntejos@das.uchile.cl}
\altaffiltext{2}{Department of Astronomy and Astrophysics, 
UCO/Lick Observatory;
University of California, 1156 High Street, 
Santa Cruz, CA 95064; xavier@ucolick.org}
\altaffiltext{3}{Department of Astronomy; University of Chicago;
5640 S. Ellis Ave., Chicago, IL 60637; hchen@oddjob.uchicago.edu}
\altaffiltext{4}{Observatoire de Gen\`eve, 51 Ch. des Maillettes, 
1290 Sauverny, Switzerland}

%% Notice that each of these authors has alternate affiliations, which
%% are identified by the \altaffilmrk after each name.  Specify alternate
%% affiliation information with \altaffiltext, with one command per each
%% affiliation.

%% Mark off your abstract in the ``abstract'' environment. In the manuscript
%% style, abstract will output a Received/Accepted line after the
%% title and affiliation information. No date will appear since the author
%% does not have this information. The dates will be filled in by the
%% editorial office after submission.

%JXP -- Do a spell check on the full document

\begin{abstract}
We report on the statistics of strong 
($W_r > 0.15$ \ \AA) \ion{C}{4} absorbers at $z=1.5-3.5$
toward high-redshift
gamma-ray bursts (GRBs). In contrast with a recent survey 
for strong \ion{Mg}{2} absorption systems at $z < 2$, 
we find that the number of \ion{C}{4}
absorbers per unit redshift $dN/dz$ does not show a significant deviation
from previous surveys using quasi-stellar objects (QSOs) as background sources. We find that the number density of \ion{C}{4} toward GRBs is 
$dN/dz|_{\rm GRB} (z \sim 1.5)= 2.2_{-1.4}^{+2.8}$, 
$dN/dz|_{\rm GRB} (z \sim 2.5)= 2.3_{-1.1}^{+1.8}$ and 
$dN/dz|_{\rm GRB} (z \sim 3.5)= 1.1_{-0.9}^{+2.6}$. 
These numbers are consistent with previous \ion{C}{4} surveys 
using QSO spectra. 
Binning the entire dataset, we set a 95$\%$ c.l.\ upper limit
to the excess of \ion{C}{4} absorbers along GRB sightlines at
twice the incidence observed along QSO sightlines.
Furthermore, the distribution of equivalent widths of the GRB
and QSO samples are consistent with being drawn from the
same parent population.
Although the results for \ion{Mg}{2} and \ion{C}{4} absorbers 
along GRB sightlines appear to contradict one another, we note
that the surveys are nearly disjoint: the \ion{C}{4} survey
corresponds to higher redshift and more highly ionized gas than the
\ion{Mg}{2} survey. Nevertheless, analysis on 
larger statistical samples may constrain properties of the galaxies
hosting these metals (e.g.\ mass, dust content)
and/or the coherence-length of the gas giving rise to the
metal-line absorption.
\end{abstract}

%% Keywords should appear after the \end{abstract} command. The uncommented
%% example has been keyed in ApJ style. See the instructions to authors
%% for the journal to which you are submitting your paper to determine
%% what keyword punctuation is appropriate.


%% From the front matter, we move on to the body of the paper.
%% In the first two sections, notice the use of the natbib \citep
%% and \citet commands to identify citations.  The citations are
%% tied to the reference list via symbolic KEYs. The KEY corresponds
%% to the KEY in the \bibitem in the reference list below. We have
%% chosen the first three characters of the first author's name plus
%% the last two numeral of the year of publication as our KEY for
%% each reference.

%% Authors who wish to have the most important objects in their paper
%% linked in the electronic edition to a data center may do so by tagging
%% their objects with \objectname{} or \object{}.  Each macro takes the
%% object name as its required argument. The optional, square-bracket 
%% argument should be used in cases where the data center identification
%% differs from what is to be printed in the paper.  The text appearing 
%% in curly braces is what will appear in print in the published paper. 
%% If the object name is recognized by the data centers, it will be linked
%% in the electronic edition to the object data available at the data centers  
%%
%% Note that for sources with brackets in their names, e.g. [WEG2004] 14h-090,
%% the brackets must be escaped with backslashes when used in the first
%% square-bracket argument, for instance, \object[\[WEG2004\] 14h-090]{90}).
%%  Otherwise, LaTeX will issue an error. 

\section{Introduction}
After extensive studies on absorption line systems in spectra
of quasi-stellar objects (QSOs) it has been widely accepted that the
majority of systems discovered are intervening and that they trace the
cosmological expansion of the universe. Various surveys have been
conducted that focus on the cosmological evolution of metal absorption
systems for distinct atomic transitions. A standard observational
measure is the number of absorbers per unit redshift, $dN/dz$, which
includes the cosmological evolution in the physical pathlength and
evolution intrinsic to the absorbers themselves
\citep[e.g.,][]{lanzetta1987,sargent1988,steidel1990}.

Since the production of heavy elements takes place in the stars which
mainly group into galaxies, it is natural to relate metal absorbers to
galaxies \citep[although this has proven difficult to test, especially
at high redshift; e.g.,][]{churchill2004, tripp05}. Therefore, studying
the evolution of metal absorbers may impact our understanding of both
galaxy evolution and the physics of the intergalactic medium (IGM).

Recently, \citet{prochter2006} have found a significant overabundance
(a factor of $\sim 4$) of strong (rest-frame equivalent width of 
\ion{Mg}{2}~2796, $W_{r}(2796) \ge 1.0$ \r A) \ion{Mg}{2}
absorbers in lines-of-sight toward gamma-ray bursts (GRBs) when
compared to the statistics drawn from lines-of-sight toward QSOs. This
result is striking because a key hypothesis of this experiment is that
intervening absorbers are independent of the background source.
Several physical effects have been proposed to explain the
overabundance: 
(1) dust within the \ion{Mg}{2} absorbers may obscure
faint QSOs, 
(2) the \ion{Mg}{2} gas may be intrinsic to the GRBs, 
(3) the GRBs may be gravitationally lensed by these absorbers, and 
(4) the absorbers are small enough that different `beam sizes' between GRBs
and QSOs may affect the statistics
\citep{prochter2006,frank2006,hao06}. \citet{prochter2006} and  
\citet{porciani07} have argued that none of these explanations is
likely to explain the full effect but it may be possible for several
to contribute together to resolve the discrepancy.

In this paper, we revisit high-resolution GRBs spectra and look for
\ion{C}{4} absorbers to obtain the first statistics of such absorbers
in this type of lines-of-sight. Because the C$^{+3}$ ion has a much
higher ionization potential than Mg$^+$, the \ion{C}{4} doublet is
likely to trace more diffuse and hotter gas. Therefore, \ion{C}{4}
systems may represent a different "population" of larger "cross
section" absorbers and the study of their statistics may help address
the problem opened by the \citet{prochter2006} result.

The paper is organized as follows. In $\S$~\ref{data} we present the
spectra used in this study. The statistical
sample and the redshift number density calculation are
described in $\S$~\ref{CIV}. The results are summarized in
$\S$~\ref{results} and discussed in $\S$~\ref{discuss}.

\section{Data}\label{data}
We have drawn our spectral sample from echelle observations summarized
in Table \ref{table1}. We have restricted our analysis to the
sightlines with at least coverage redward of the \lya\ forest
(i.e. $z_{GRB}^{Ly\alpha} < z^{CIV} \le z_{GRB}^{CIV}$) and to spectra
with moderate to high signal-to-noise ratio ($>5$\,pix$^{-1}$). The
sample is comprised of echelle spectra of the following GRB
afterglows: GRB021004, GRB050730, GRB050820, GRB050922C and GRB060607.
The data was obtained with the MIKE/Las Campanas \citep{mike},
UVES/VLT \citep{uves} and HIRES/Keck \citep{vogt94} echelle
spectrographs allowing high spectral resolution of $R \sim 40000$. The
UVES data were retrieved from the ESO archive and processed with the
standard UVES pipeline. For each GRB the individual spectra
acquired with a given instrument were reduced, coadded and normalized
with standard techniques. We refer the reader to the following papers
for a full discussion of the observations
\citep{fiore05,chen05,prochaska07,fiore07,ledoux06}.

\section{\ion{C}{4} Absorption Systems toward GRBs}\label{CIV}

\subsection{Equivalent Widths}
Equivalent widths were calculated using both direct pixel integration,
$W^{pixel}$, and a Gaussian fit, $W^{gauss}$, applied directly to the
high-resolution data (since previous surveys used spectra of lower
resolution, we first confirmed that smoothing and re-binning our data
does not alter the equivalent widths). We obtained $W^{pixel}$ over
the smallest velocity window that contain the whole system. The rest
frame equivalent width $W_r$ was calculated by taking
$W_r=W_{obs}/(1+z)$. We arbitrarily defined the redshift of the
absorption system such that it appears centered in velocity space.

\begin{figure}
\plotone{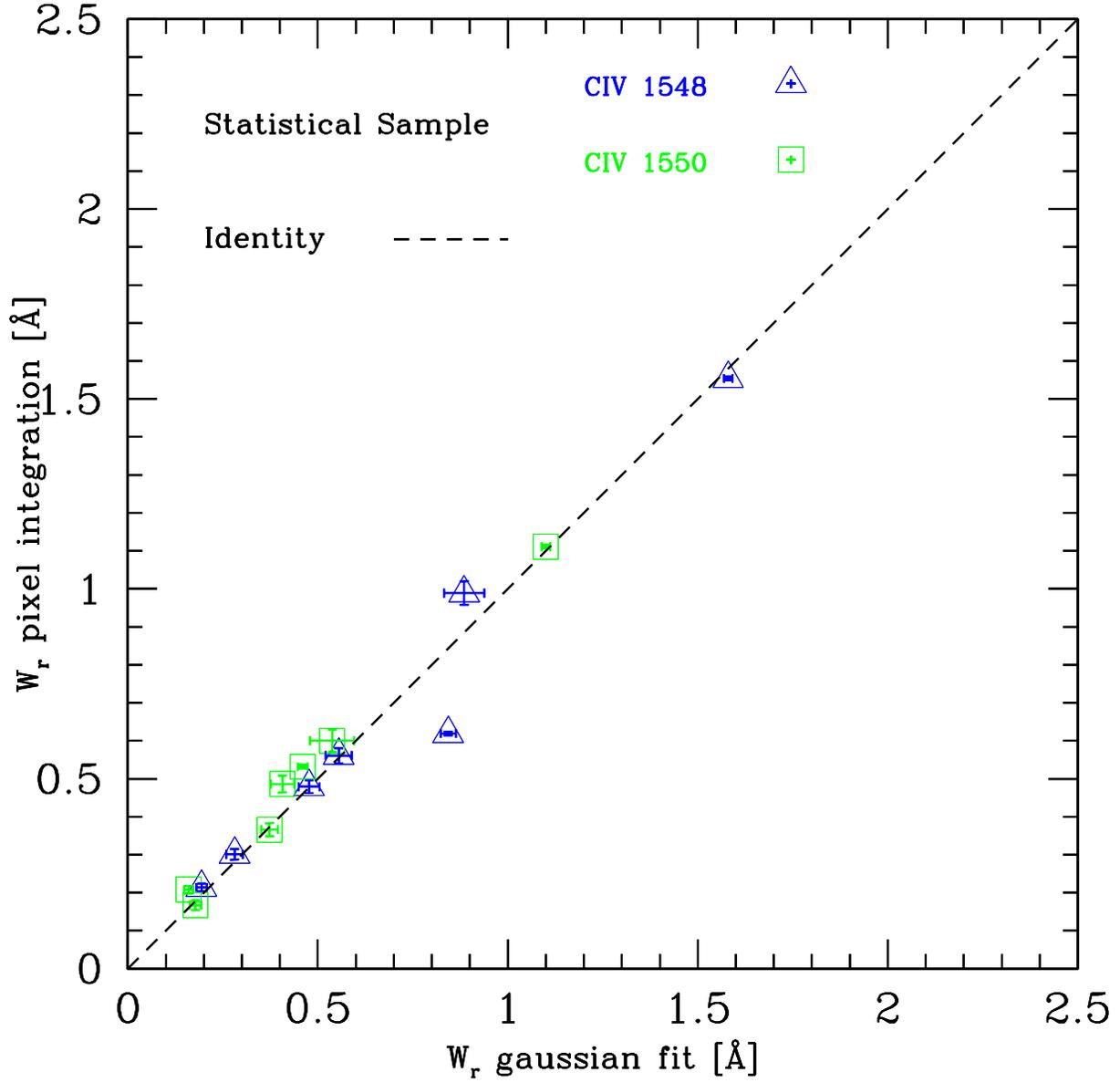}
\caption{Rest frame equivalent width of \ion{C}{4} in the Full Sample measured with Gaussian fit and with pixel integration. The dashed line is the identity function.}\label{Ws}
\end{figure}

\begin{figure}
\plotone{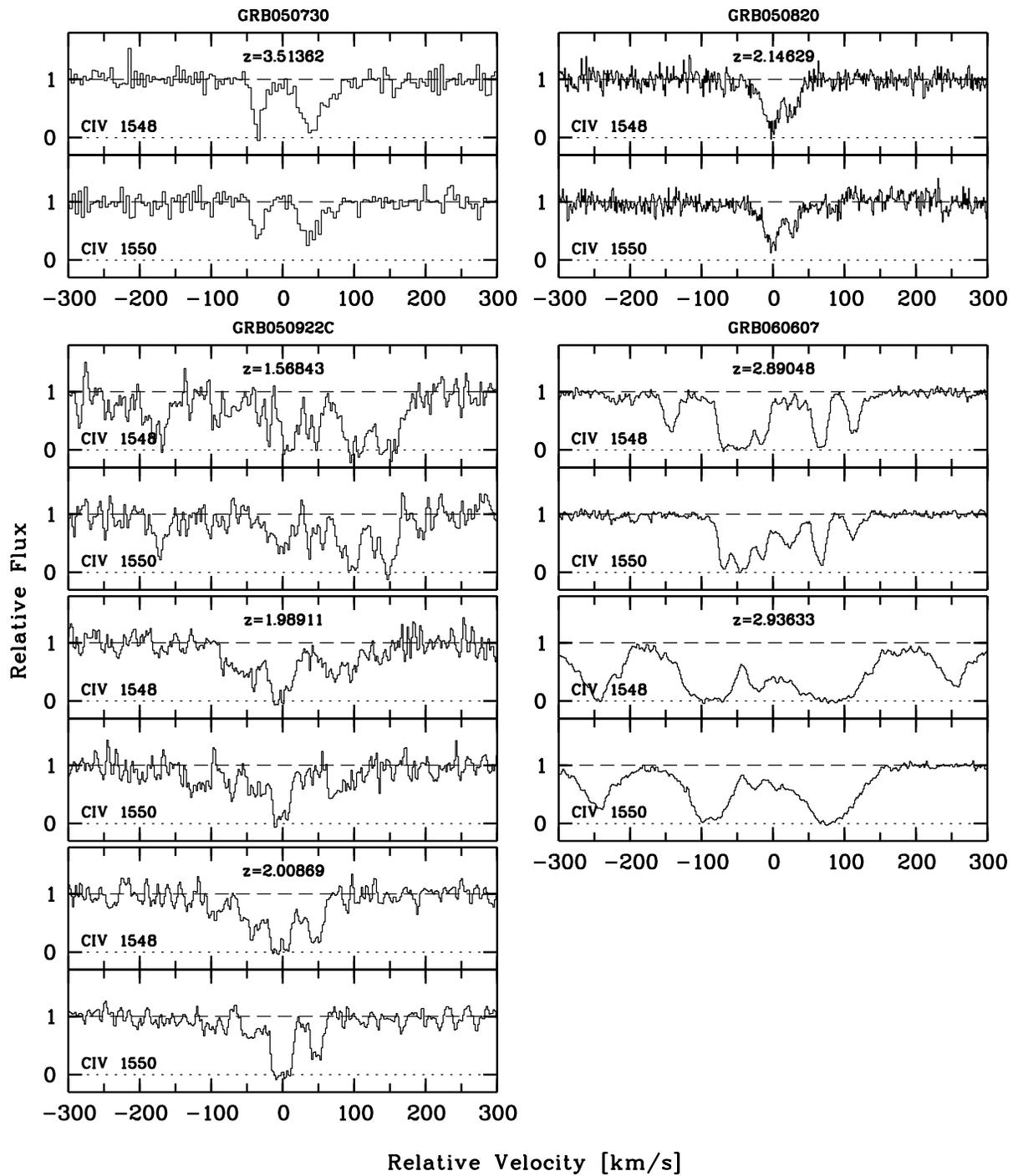}
\caption{Velocity profiles of \ion{C}{4} absorption systems toward GRBs in the Statistical Sample.}\label{figint}
\end{figure}

To obtain $W^{gauss}$, we fitted a single or multiple Gaussian profile
to the absorption lines, and summed each component of an individual
system. In the case of multiples components, the error associated is,
$\sigma_{system}^2=\sum_i \sigma_{i}^2$, where $\sigma_{i}$ is the
individual error of component $i$. Figure~\ref{Ws} shows that
equivalent widths from both methods are, in general,
self-consistent. The outlier is due to the blended feature in
the main component of the system at $z=2.89048$ (see
Figure~\ref{figint}). Hereafter, we will use only $W^{gauss}$ because
of its robustness (continuum and blended systems are fitted,
independently of velocity window).

\subsection{Definition of the Samples}\label{sample}
\subsubsection{Full Sample}
We first define a Full Sample of \ion{C}{4} absorption systems that
satisfy the following criteria: (1) $z_{GRB}^{Ly\alpha}<z_{abs}^{CIV}
\le z_{GRB}^{CIV}$; i.e., the Full Sample includes lines associated
with the GRB host galaxy (we will later restrict the analysis to
intervening \ion{C}{4} systems). (2) The equivalent width of the
\ion{C}{4} doublet must be detected at the 5$\sigma$ level or
higher. No conditions were imposed to the equivalent width ratio for
both components although all members of the Statistical Sample
(see below) exhibit an equivalent width ratio $W(1548)/W(1550)$
between 1 (saturated regime) and 2 (unsaturated regime). (3)
A complex system is considered a single system if the velocity
components group within $500$ \kms. This last condition is useful to
compare with previous surveys that were made at lower spectral
resolution \citep[for instance,][used a velocity window of $150$ \kms\
to group components of a single system but then defined a statistical
sample counting as a single system all components within $1000$ \kms;
Our Statistical Sample, defined below, does not have systems separated
by less than $1000$ \kms, therefore these two samples are
comparable]{steidel1990}.

Although performed redward of the \lya\ forest, the \ion{C}{4} survey
has one significant contaminant: the \ion{O}{1}~1302/\ion{Si}{2}~1304
pair of transitions has nearly identical separation as the \ion{C}{4}
doublet. It is trivial, however, to identify these contaminants by
searching for other \ion{Si}{2} transitions and the \ion{C}{2}~1334
transition.

We found 29 candidate \ion{C}{4} absorption systems in our
Full Sample. This number includes absorption systems with $z \sim
z_{GRB}$. Table \ref{table2} shows the information on redshift limits
of each line-of-sight, $z_{abs}^{CIV}$ and $W_r$.
 
\subsubsection{Statistical Sample}
To compare the statistics of \ion{C}{4} absorption along GRB
sightlines with that one for QSO sightlines, one must define a sample
with identical equivalent width limits. Previous surveys have defined
this limit to $W_{min} = 0.15$ \AA\ in the rest frame
\citep{steidel1990,misawa2002}. We adopt the same value, i.e., $W_r >
0.15$ \AA \ in \textit{both} members of the \ion{C}{4} doublet.

In addition, to avoid contributions from galaxies clustered with the
hosts of GRB and wind features that might be associated with the GRB
progenitor \citep{chen07} and to compare with previous surveys,
we define a Statistical Sample with the same $z_{start}$ as in
the Full Sample but $z_{end}$ at $5000 \ km \ s^{-1}$ from $z_{GRB}$
\citep[as in][for $z_{QSO}$]{steidel1990}.

All the spectra included in our Statistical Sample permit detections
of lines with $W_r > 0.15$ \AA\ with at least $5\sigma$ significance
from $z_{GRB}^{Ly\alpha}$ to $5000 \ km \ s^{-1}$ within
$z^{CIV}_{GRB}$. The Statistical Sample is composed of 7 absorption
systems (see systems labeled with a dagger in Table \ref{table2} and
see Figure \ref{figint} for velocity plots). We excluded 2
systems due to their dubious character (systems labeled with an 'X' in
Table \ref{table2}): $z_{abs}=3.254$, although not saturated, it shows
an equivalent width ratio $<  1$; $z_{abs}=2.3589$ falls very
close to $W_{min}$ \AA\ but $W^{gauss}_r(1550) < 0.15$ \AA.

\subsection{Redshift Number Density}\label{dNdz}
\begin{figure}
\plotone{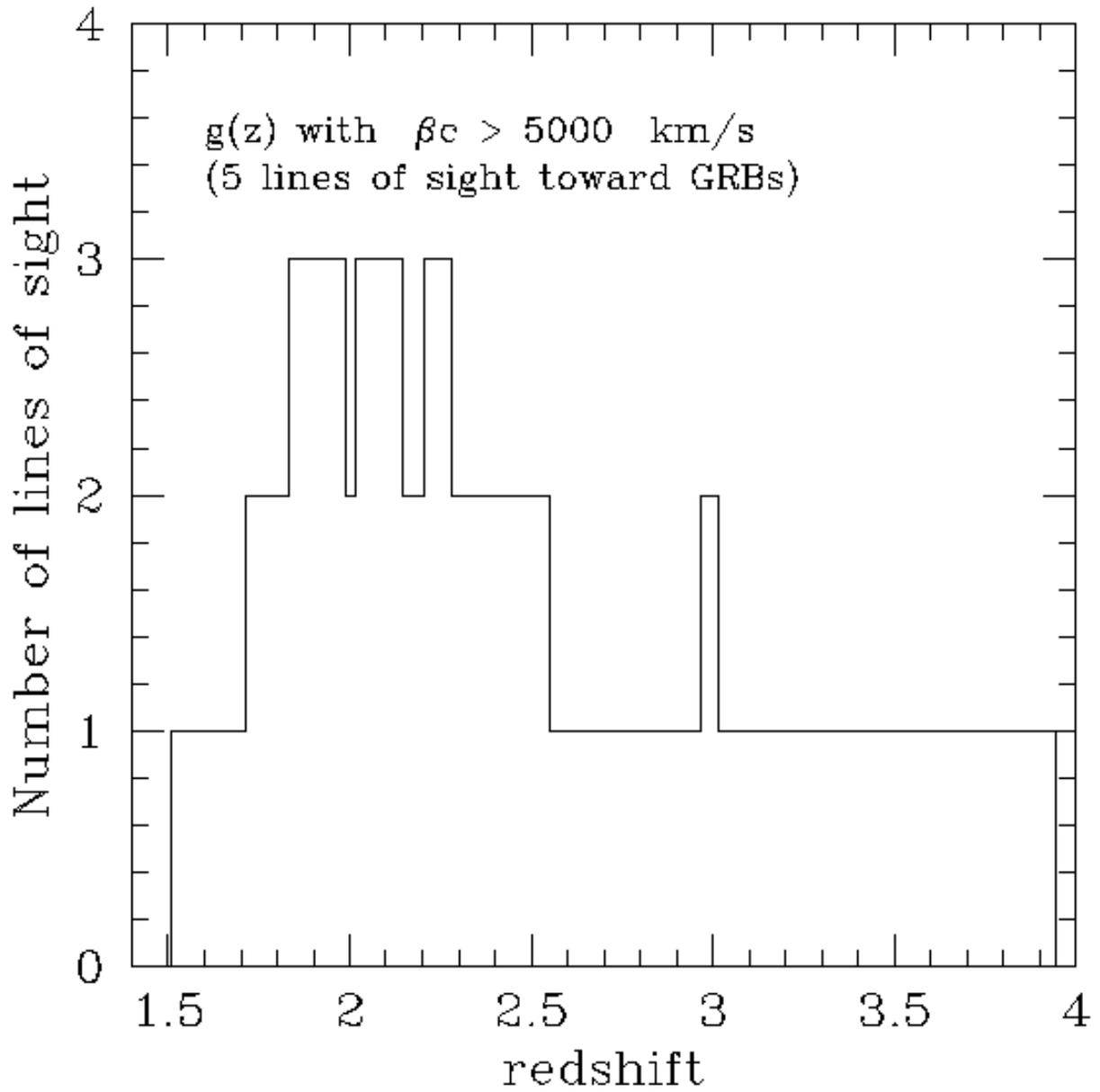}
\caption{Number of GRB lines-of-sight available for a \ion{C}{4} survey
with $W_r > 0.15$\r A\ using $\beta c=5000$ \kms\ 
to discount the contamination of the source.}\label{gz_b5}
\end{figure}

\begin{figure}[h]
\begin{center}
\plotone{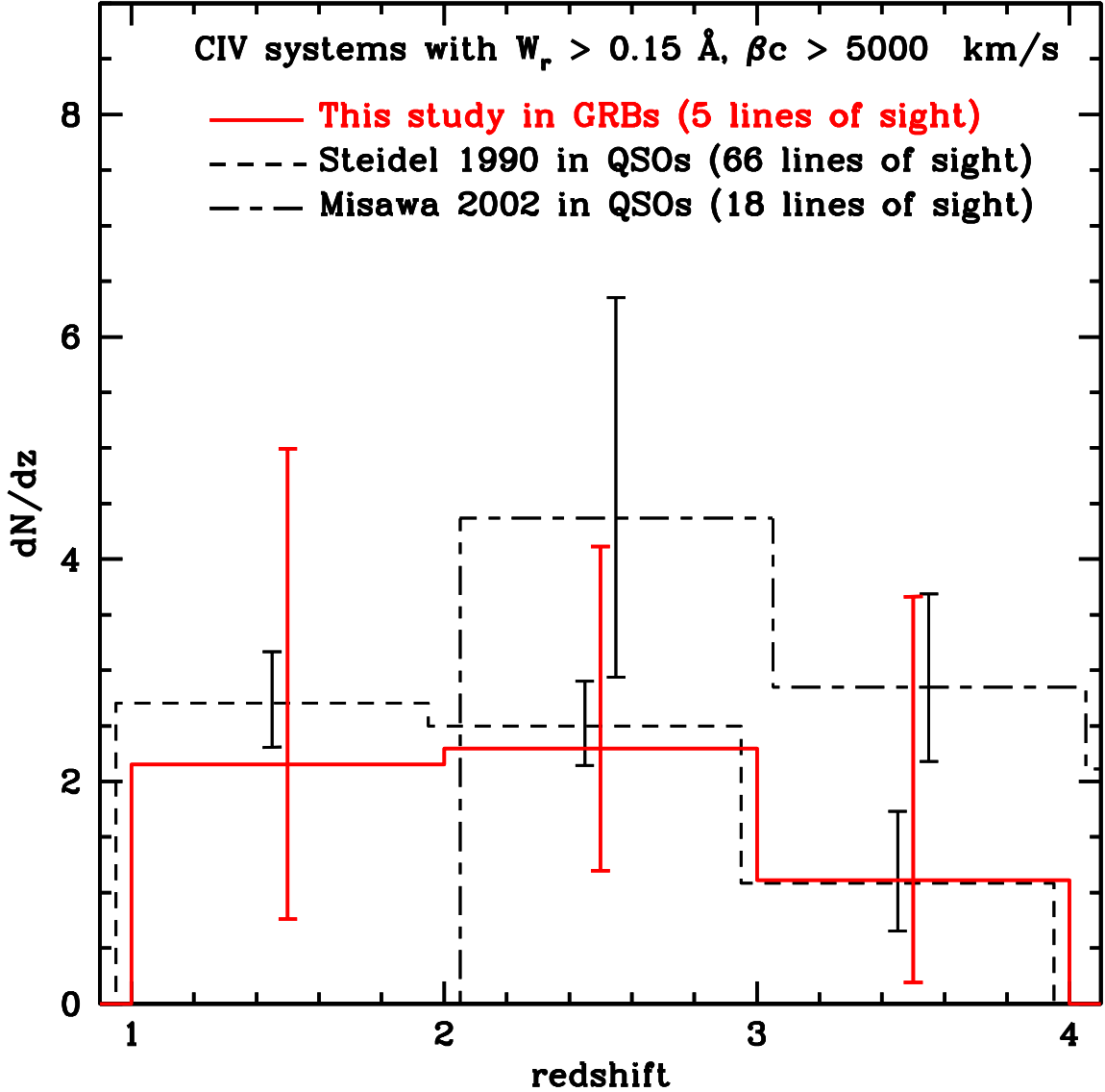}
\caption{Distribution of the number of \ion{C}{4} absorption systems
per unit redshift in our GRB Statistical Sample. As a comparison
(dashed line) we show results from QSO survey by \citet{steidel1990}
and \citet{misawa2002} using the same binning as that for the
GRB-\ion{C}{4} analysis. The bins have been slightly offset in
redshift for clarity.}
\label{fig_comp2}
\end{center}
\end{figure}

The redshift number density, $dN/dz$, is the number of absorbers per
unit redshift. By definition, $dN/dz = N_{abs}/ \Delta z$ where
$N_{abs}$ is the number of absorption systems in the redshift path
$\Delta z$. Defining the number of lines-of-sight between $z$ and $z$
$+ dz$ as $g(z)$ (a function that defines the redshift path density),
the redshift path can be written as $\Delta z = \sum g(z) \delta z$
where $\delta z$ is the grid resolution. Figure \ref{gz_b5} shows
$g(z)$ for our sample. Figure \ref{fig_comp2} shows our result on
$dN/dz$ toward GRBs ($dN/dz|_{\rm GRB}$). The choice of redshift bins
is arbitrary; we used 3 bins of 1 redshift unit between $z=1$ and
$z=4$.

 To estimate the error in $dN/dz$ we assume a Poissonian distribution
in the number of absorbers $N_{abs}$ in a given redshift bin $\Delta
z^{bin}$ and then assign the upper and lower limit corresponding to
$1\sigma$ confidence level of a Gaussian distribution, ($1\sigma =
0.8413$) $1\sigma_{N_{abs}}^+$ and $1\sigma_{N_{abs}}^-$ respectively
obtained from \citet{gehrels86}. Therefore, $\sigma_{dN/dz}^{+,-} =
1\sigma_{N_{abs}}^{+,-}/\Delta z^{bin}$. This is done for each
redshift bin.

\section{Result} \label{results}
For the Statistical Sample we find $dN/dz|_{\rm GRB}(z \sim
1.5)= 2.2_{-1.4}^{+2.8}$, $dN/dz|_{\rm GRB}(z \sim 2.5)=
2.3_{-1.1}^{+1.8}$ and $dN/dz|_{\rm GRB}(z \sim 3.5)=
1.1_{-0.9}^{+2.6}$. Table \ref{table3} shows the specific results for
each redshift bin.

How do these numbers compare with $dN/dz|_{\rm QSO}$? One of the
largest published surveys of \ion{C}{4} along QSO sightlines was
carried out by \citet{steidel1990}. In that survey, 66 QSO spectra
with low spectral resolution ($\sim 1-2$ \AA) were used (55 of them
were obtained from \citet{sargent1988}). We used their result as
comparison. \citet{steidel1990} measured the incidence of \ion{C}{4}
systems with $W_{min} = 0.15$ \r A and parameterized their results as
$dN/dz|_{\rm QSO} = N_0 (1+z)^{-1.26}$. However, in what follows, we
compute $dN/dz|_{\rm QSO}$ directly from the tables in
\citet{steidel1990} and \citet{sargent1988}. We also compare with
another published survey by \citet{misawa2002} (18 QSO spectra with
resolution of $\sim 2$ \AA) although this survey has much lower
redshift coverage and does not provide a better sensitivity ($W_{min}
= 0.15$ \r A). 

We find that $dN/dz|_{\rm GRB}$ decreases with redshift in the range
between $z=1$ and $z=4$ as in QSO surveys. In Figure \ref{fig_comp2}
we compare $dN/dz|_{\rm GRB}$ with $dN/dz|_{\rm QSO}$ using our
binning. It can be seen that, within errors, our result on \ion{C}{4}
matches the \citet{steidel1990} and the \citet{misawa2002} ones
(according to Misawa \textit{et al.}, the fact that these two previous
results do not match each other would be due to an statistical
accident). On the other hand, if we use only one redshift bin between
$z=1$ and $z=4$ we find that the incidence of \ion{C}{4} absorbers
along GRB sightlines is less than twice that one along QSOs (Steidel
1990 sample) at the 95$\%$ confidence level, i.e.,\ a large
overabundance is improbable (even taking into account the 2 dubious
systems excluded from the Statistical Sample). Therefore, there is no
significant difference between the GRB and QSO statistics in our
sample.

\section{Discussion}\label{discuss}

The principal result of our analysis is that the incidence of
\ion{C}{4} absorbers ($W_r \ge 0.15$\r A) along GRB sightlines is
consistent with that observed along QSO sightlines. Note that
the same conclusion is reached by \citet{sudilovsky2007} using similar
data (although these authors use a smaller total redshift path and a
lower equivalent width cutoff than ours). 

Quantitatively, we set a 95$\%$ upper limit of $\mdnzcg < 2 \mdnzcq$
over the redshift interval $z \sim 2$ to 4. At the surface, this
result lies in stark contrast to the incidence of strong \ion{Mg}{2}
absorbers at $z<2$ where \citet{prochter2006} find a {\it lower limit}
$\mdnzmgg > 2 \mdnzmgq$ at the 99.9$\%$ significance level.

Are these two results contradictory? Let us first stress that the two
absorption-line samples may be nearly disjoint. Firstly, one important
difference is that the \ion{Mg}{2} absorbers were surveyed at $z<2$
while most of the pathlength surveyed here for \ion{C}{4} absorption
has $z>2$. While there is no indication that the \dnzmgg\ enhancement
is declining with increasing redshift, there are possible reasons to
expect such an effect (e.g.\ gravitational lensing efficiency is
maximal at $z_{lens} \approx z_{GRB}/2$). In that case the \ion{Mg}{2}
enhancement would vanish at the redshifts probed here with \ion{C}{4}.

But even if this were not the case, let us consider a second argument
in favor of the present results. The key issue is that \ion{C}{4} may
track a distinct phase of gas from \ion{Mg}{2} in the intergalactic
medium owing to its substantially higher ionization potential. This is
reflected in the statistics of QSO surveys. At $z=2$, $\mdnzmgq \approx
0.4$ \citep{prochter2006a} and $\mdnzcq \approx 2.5$
\citep{steidel1990}, i.e., \ion{C}{4} systems are a factor of 6 more
abundant than \ion{Mg}{2} for the $W_r$ thresholds relevant to our
study. In other words, there is plenty of \ion{C}{4} (weak) systems
that do not show strong \ion{Mg}{2} absorption.

On the other hand, we expect that nearly every strong \ion{Mg}{2}
absorber will also exhibit \ion{C}{4} absorption.  Furthermore, many
of these will have an equivalent width in excess of 0.15\AA\
\citep[e.g.][]{churchill1999}. Therefore, if the enhancement in
\dnzmgg\ continues beyond $z=2$, we do expect a certain bias to larger
\dnzcg. However, since there is not a one-to-one correspondence
between $W_r \ge 1$\AA\ \ion{Mg}{2} absorbers and $W_r \ge 0.15$\AA\
\ion{C}{4} absorbers, such possible bias is diluted in the much more
numerous statistics of \ion{C}{4} (indeed, none of the \ion{C}{4}
absorbers in our sample that have coverage of the \ion{Mg}{2} doublet
shows significant absorption; e.g., $z_{CIV} = 1.568, 1.989$ from
GRB~050922C). Even if we assume that $\mdnzmgg = 4~ \mdnzmgq$ and that
each system also has $W_{CIV} > 0.15$\AA\, the effect on \dnzcg\ is
less than $50\%$ (at most (\dnzcg)/(\dnzcq) = 9/6 given that
(\dnzmgg)/(\dnzmgq) = 4/1 and (\dnzcq)/(\dnzmgq) = 6/1).

\begin{figure}[h]
\begin{center}
\plotone{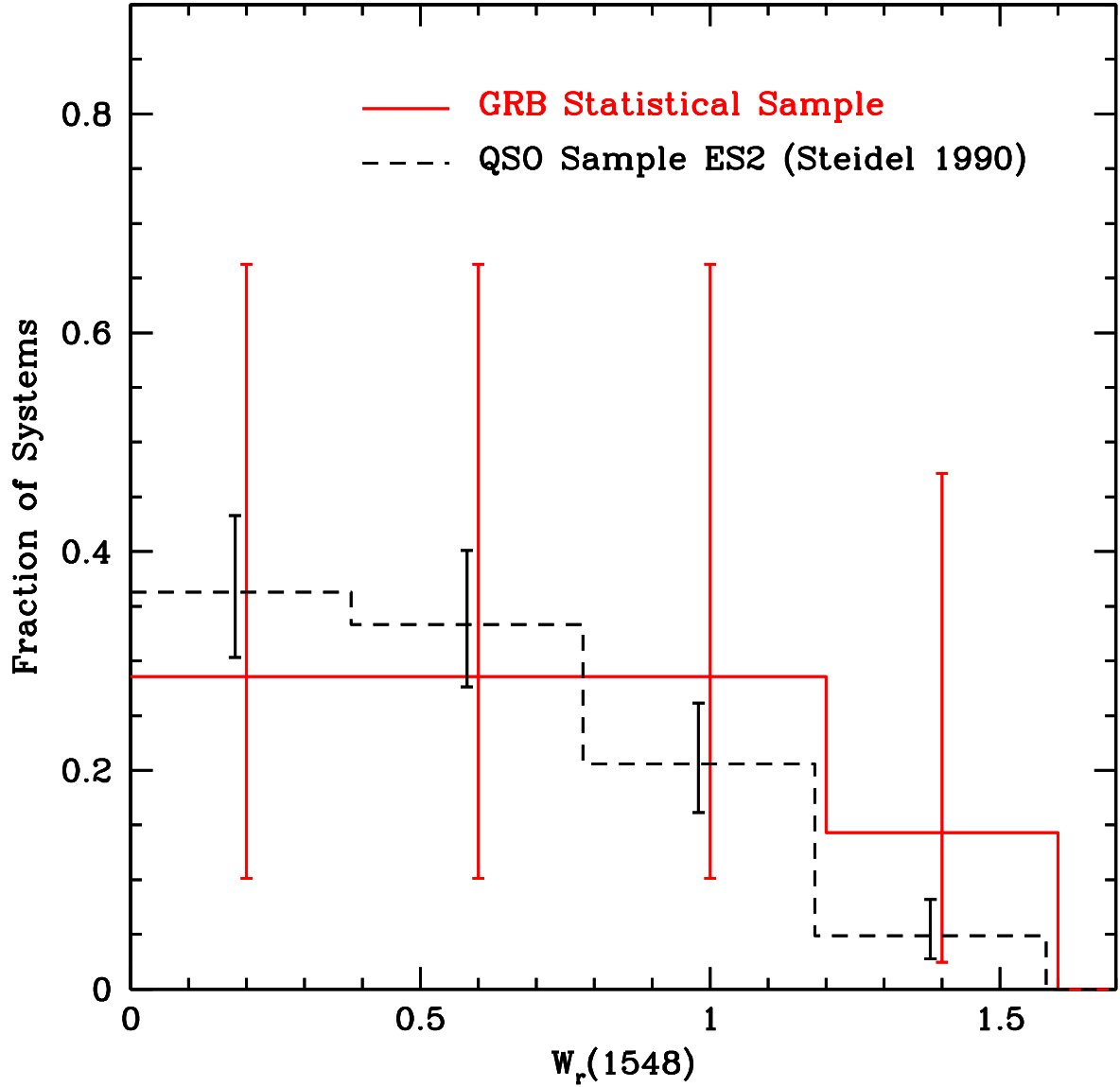}
\caption{Distribution of the rest frame equivalent widths $W_r(1548)$
for our Statistical Sample compared with the \citet{steidel1990} one (dashed line). The bins have been slightly offset in $W_r(1548)$ for clarity.}
\label{fig_Wdist}
\end{center}

\end{figure}

While the change to \dnzcg\ for $W_{1548} > 0.15$\AA\ may be modest,
as argued above, it is possible that an enhancement in \dnzmgg\ will
imply an \ion{C}{4} equivalent width {\it distribution} that is very
different from that toward QSOs, e.g., a much higher incidence of
$W_r(1548) > 1$\AA. Figure~\ref{fig_Wdist} presents a histogram of the
equivalent widths for our analysis and that of \citet{steidel1990} with
arbitrary binning. We performed a K-S test on the un-binned
distributions of rest frame equivalent widths that rules out the null
hypothesis (of similar distributions) at only the 1\% c.l. Therefore,
there is no indication of possible differences between equivalent
width distributions of \ion{C}{4} absorbers toward GRBs and QSOs.

In summary, there is no fundamental conflict between observing an
enhanced incidence of strong \ion{Mg}{2} absorbers along GRB
sightlines without a corresponding enhancement of \ion{C}{4}
absorbers. Before concluding, let us speculate on the implications of
these results under the assumption that future data confirm the
currently measured incidences. If our results on \dnzcg\ and the
current enhancement in \dnzmgg\ are confirmed in larger surveys, one
may use these contrasting results to infer characteristics of the
\ion{C}{4} absorbers. For example, if the \dnzmgg\ enhancement is to
be explained by the geometric differences between the GRB and
QSO beams (comparable with the characteristic size of \ion{Mg}{2}
`clouds') \citep{frank2006}, then one would conclude \ion{C}{4}
absorbers are significantly larger than \ion{Mg}{2} clouds and the QSO
beam size. Indeed, such differences have been inferred from
comparisons of metal-line absorption along gravitationally lensed
QSOs \citep[e.g.,][]{lopez99}. 

Similarly, the results could constrain the dust redenning or
mass of the galaxies hosting \ion{C}{4} absorbers. Correlations
between these quantities and \ion{Mg}{2} equivalent width have been
found by recent studies using \ion{Mg}{2} statistics derived from
Sloan Digital Sky Survey QSOs. For instance, \citet{menard2007} have
parameterized the (significant) amount of QSO redenning due to
\ion{Mg}{2} absorbers, and \citet{bouche2006} have found an
anti-correlation between \ion{Mg}{2} equivalent width and mass of the
dark-matter halo hosting the absorbing galaxy.  The different behavior
of \ion{C}{4} and \ion{Mg}{2}, as revealed by the statistics toward
GRBs, then can put constraints on these absorber properties by
studying \ion{Mg}{2} samples selected by \ion{C}{4}.

%% To help institutions obtain information on the effectiveness of their
%% telescopes, the AAS Journals has created a group of keywords for telescope
%% facilities. A common set of keywords will make these types of searches
%% significantly easier and more accurate. In addition, they will also be
%% useful in linking papers together which utilize the same telescopes
%% within the framework of the National Virtual Observatory.
%% See the AASTeX Web site at http://www.journals.uchicago.edu/AAS/AASTeX
%% for information on obtaining the facility keywords.

%% After the acknowledgments section, use the following syntax and the
%% \facility{} macro to list the keywords of facilities used in the research
%% for the paper.  Each keyword will be checked against the master list during
%% copy editing.  Individual instruments or configurations can be provided 
%% in parentheses, after the keyword, but they will not be verified.

%{\it Facilities:} \facility{Nickel}, \facility{HST (STIS)}, \facility{CXO (ASIS)}.

%% Appendix material should be preceded with a single \appendix command.
%% There should be a \section command for each appendix. Mark appendix
%% subsections with the same markup you use in the main body of the paper.

%% Each Appendix (indicated with \section) will be lettered A, B, C, etc.
%% The equation counter will reset when it encounters the \appendix
%% command and will number appendix equations (A1), (A2), etc.

\clearpage

%% Use the figure environment and \plotone or \plottwo to include
%% figures and captions in your electronic submission.
%% To embed the sample graphics in
%% the file, uncomment the \plotone, \plottwo, and
%% \includegraphics commands
%%
%% If you need a layout that cannot be achieved with \plotone or
%% \plottwo, you can invoke the graphicx package directly with the
%% \includegraphics command or use \plotfiddle. For more information,
%% please see the tutorial on "Using Electronic Art with AASTeX" in the
%% documentation section at the AASTeX Web site,
%% http://www.journals.uchicago.edu/AAS/AASTeX.
%%
%% The examples below also include sample markup for submission of
%% supplemental electronic materials. As always, be sure to check
%% the instructions to authors for the journal you are submitting to
%% for specific submissions guidelines as they vary from
%% journal to journal.

%% This example uses \plotone to include an EPS file scaled to
%% 80% of its natural size with \epsscale. Its caption
%% has been written to indicate that additional figure parts will be
%% available in the electronic journal.

\clearpage

\begin{table}[h]
\begin{center}
\begin{tabular}{lclc}
\tableline \tableline
GRB    & $z_{GRB}$ & Instrument & Reference\\
\tableline          
021004 &  2.335  &  UVES & 1\\      
050730 &  3.97   &  MIKE&2,3 \\ 
050820 &  2.6147 &  HIRES &3 \\
050922C&  2.199  &  UVES& 4\\   
060607 &  3.082  &  UVES& 5\\
\tableline
\end{tabular}
\caption{Spectroscopic observations of our GRB sample. References: (1)
\citet{fiore05}; (2) \citet{chen05}; (3) \citet{prochaska07}; (4)
\citet{fiore07}; (5) \citet{ledoux06}} \label{table1}
\end{center}
\end{table}

%%
%% End of file `sample.tex'.

\clearpage

\begin{table}[h]
\scriptsize
\begin{center}
\begin{tabular}{lcccllcccc}
\tableline\tableline
GRB   & $z_{GRB}$ & $z_{start} $& $z_{end}^{\beta c=5000 km/s} $&$z_{abs}^{CIV}$&$ W_r^{pixel}(1548) $&$ W_r^{pixel}(1550) $&$ W_r^{gauss}(1548) $&$ W_r^{gauss}(1550) $\\
\tableline
021004  & 2.335  &1.715$^a$  &2.279$^a$  &1.74738&          0.035 $\pm$ 0.006& 0.039 $\pm$ 0.007& 0.031 $\pm$ 0.006 & 0.051 $\pm$ 0.014\\
 &       &       &       &1.81108&                          0.158 $\pm$ 0.011& 0.061 $\pm$ 0.010& 0.191 $\pm$ 0.014 & 0.073 $\pm$ 0.007 \\
 &       &       &       &1.83415&                          0.153 $\pm$ 0.008& 0.051 $\pm$ 0.007& 0.141 $\pm$ 0.012 & 0.044 $\pm$ 0.010\\
 &       &       &       &2.29800&                          0.898 $\pm$ 0.011& 0.561 $\pm$ 0.012& 0.898 $\pm$ 0.018 & 0.520 $\pm$ 0.016  \\
 &       &       &       &2.32800&                          2.094 $\pm$ 0.012& 1.695 $\pm$ 0.012& 2.584 $\pm$ 0.045 & 1.786 $\pm$ 0.021   \\
050730  & 3.97   &2.965  &3.887  &3.25400  $\times$&        0.233 $\pm$ 0.014& 0.266 $\pm$ 0.013& 0.249 $\pm$ 0.023 & 0.300 $\pm$ 0.023\\
 &       &       &       &3.51362         $ \dagger$  &     0.301 $\pm$ 0.014& 0.167 $\pm$ 0.013& 0.281 $\pm$ 0.022 & 0.178 $\pm$ 0.015 \\
 &       &       &       &3.63952&                          0.063 $\pm$ 0.007& 0.072 $\pm$ 0.007& 0.059 $\pm$ 0.009 & 0.062 $\pm$ 0.009 \\
 &       &       &       &3.96803&                          0.811 $\pm$ 0.017& 0.692 $\pm$ 0.019& 0.874 $\pm$ 0.026 & 0.754 $\pm$ 0.027   \\
050820  & 2.6147&1.833   &2.554  &1.94010&                  0.060 $\pm$ 0.005& 0.122 $\pm$ 0.004& 0.068 $\pm$ 0.007 & 0.137 $\pm$ 0.009\\
 &       &       &       &2.05973&                          0.041 $\pm$ 0.003& 0.058 $\pm$ 0.003& 0.039 $\pm$ 0.004 & 0.057 $\pm$ 0.006\\
 &       &       &       &2.07491&                          0.160 $\pm$ 0.008& 0.047 $\pm$ 0.008& 0.113 $\pm$ 0.009 & 0.047 $\pm$ 0.008\\
 &       &       &       &2.14629  $  \dagger$  &           0.214 $\pm$ 0.010& 0.208 $\pm$ 0.010& 0.194 $\pm$ 0.014 & 0.160 $\pm$ 0.012  \\
 &       &       &       &2.32375&                          0.168 $\pm$ 0.006& 0.142 $\pm$ 0.006& 0.150 $\pm$ 0.009 & 0.119 $\pm$ 0.010\\
 &       &       &       &2.35890        $\times$  &        0.331 $\pm$ 0.013& 0.152 $\pm$ 0.013& 0.292 $\pm$ 0.019 & 0.143 $\pm$ 0.012\\
 &       &       &       &2.61444&                          1.504 $\pm$ 0.010& 1.091 $\pm$ 0.012& 1.544 $\pm$ 0.040 & 1.116 $\pm$ 0.106  \\
050922C &  2.199&1.511   &2.146 &1.56843  $ \dagger$&       0.989 $\pm$ 0.031& 0.600 $\pm$ 0.029& 0.885 $\pm$ 0.053 & 0.538 $\pm$ 0.058  \\
 &       &       &       &1.98911         $ \dagger$&       0.561 $\pm$ 0.020& 0.486 $\pm$ 0.022& 0.555 $\pm$ 0.034 & 0.407 $\pm$ 0.031   \\
 &       &       &       &2.00869         $ \dagger$&       0.479 $\pm$ 0.017& 0.366 $\pm$ 0.017& 0.477 $\pm$ 0.027 & 0.373 $\pm$ 0.022  \\
 &       &       &       &2.07766&                          0.173 $\pm$ 0.009& 0.142 $\pm$ 0.010& 0.185 $\pm$ 0.011 & 0.138 $\pm$ 0.015\\
 &       &       &       &2.14204&                          0.040 $\pm$ 0.003& 0.037 $\pm$ 0.003& 0.044 $\pm$ 0.004 & 0.035 $\pm$ 0.004\\
 &       &       &       &2.19973&                          0.729 $\pm$ 0.009& 0.541 $\pm$ 0.009& 0.735 $\pm$ 0.022 & 0.565 $\pm$ 0.016   \\
060607  &  3.082 &2.207$^b$  &3.014$^b$ &2.21653&           0.291 $\pm$ 0.005& 0.095 $\pm$ 0.007& 0.263 $\pm$ 0.006 & 0.106 $\pm$ 0.007\\
 &       &       &       &2.27853&                          0.164 $\pm$ 0.004& 0.093 $\pm$ 0.004& 0.171 $\pm$ 0.008 & 0.100 $\pm$ 0.008\\
 &       &       &       &2.89048         $ \dagger$&       0.619 $\pm$ 0.005& 0.532 $\pm$ 0.005& 0.843 $\pm$ 0.020  &  0.460 $\pm$ 0.013  \\
 &       &       &       &2.91659&                          0.036 $\pm$ 0.003& 0.035 $\pm$ 0.003& 0.031 $\pm$ 0.003  &  0.037 $\pm$ 0.004\\
 &       &       &       &2.93633         $ \dagger$&       1.554 $\pm$ 0.005& 1.111 $\pm$ 0.005& 1.580 $\pm$ 0.011  &  1.100 $\pm$ 0.011  \\
 &       &       &       &3.04979&                          0.271 $\pm$ 0.005& 0.245 $\pm$ 0.005& 0.233 $\pm$ 0.008  &  0.224 $\pm$ 0.035  \\
 &       &       &       &3.07488&                          0.386 $\pm$ 0.004& 0.295 $\pm$ 0.004& 0.399 $\pm$ 0.009  &  0.291 $\pm$ 0.007  \\
\tableline
\end{tabular}
\caption{Full Sample: All \ion{C}{4} absorption systems found between
$z_{GRB}^{Ly\alpha} < z^{CIV} \le z_{GRB}^{CIV}$ for each GRB spectra.
$^{a}$ The region $1.988< z < 2.014$ is excluded due to insufficient
signal to noise ratio.  $^{b}$ The region $2.616< z < 2.668$ is
excluded due to a gap in the spectrum.  $\dagger$ Absorbers in the
Statistical Sample.  $\times$ Systems that were excluded (see section $3.2.2$).}
\label{table2} \end{center} \end{table}

%% %% End of file `sample.tex'.

\clearpage

\begin{table}[h]
\begin{center}
\begin{tabular}{cccc|c}
\tableline \tableline
Redshift Bin & $N_{abs}$ & $\Delta z$ &\dnzcg&\dnzcq\\
\tableline          
$[1,2[$ &  $2$ & $0.93$   &$2.2_{-1.4}^{+2.8}$&  $2.7^{+0.5}_{-0.4}$\\
$[2,3[$ &  $4$ & $1.74$   &$2.3_{-1.1}^{+1.8}$&  $2.5^{+0.4}_{-0.4}$\\
$[3,4[$ &  $1$ & $0.90$   &$1.1_{-0.9}^{+2.6}$&  $1.1^{+0.7}_{-0.4}$\\
\tableline
$[1,4[$ &  $7$ & $3.57$   &$2.0_{-0.7}^{+1.1}$&  $2.4^{+0.3}_{-0.2}$\\
\tableline
\end{tabular}
\caption{Specifications of the resulting \dnzcg \ for each bin from
the Statistical Sample. As comparison the fifth column shows the
resulting \dnzcq \ for same binnings from \citet{steidel1990}
results.} \label{table3}
\end{center}
\end{table}

%%
%% End of file `sample.tex'.

\clearpage

\clearpage
\clearpage

\clearpage

\clearpage
\clearpage

\end{document}